# ON THE EVOLUTION OF QUASICRYSTALLINE AND CRYSTALLINE PHASES IN RAPIDLY QUENCHED Al-Co-Cu-Ni ALLOY

**T.P. Yadav\*, N.K. Mukhopadhyay[1], R.S.Tiwari and O.N.Srivastava**

Department of Physics, Banaras Hindu University, Varanasi-221005, India
[1]Department of Metallurgical Engg., IT, Banaras Hindu University, Varanasi-221005, India

*Abstract*

The occurrence of stable decagonal quasicrystalline phase in Al-Co-Ni and Al-Cu-Co alloys through conventional solidification is well established. Earlier, we have studied the effect of Cu substitution in place of Co in the $Al_{70} Co_{15}Ni_{15}$ alloy. Here we report the structural/micro-structural changes with substitution of Cu for Ni in rapidly solidified Al-Co-Ni alloys. The melt-spun ribbons have been characterized using X-ray diffractometry (XRD), Scanning and transmission electron microscopy (SEM & TEM). With an increase in Cu content in the melt spun $Al_{70} Co_{15}Cu_x Ni_{15-x}$ (x=0 to 15) alloys, the relative amount of the decagonal phase decreased up to 10 at% of Cu. At this composition the quaternary alloy showed the coexistence of decagonal quasicrystal and superstructure of $\tau_3$ vacancy ordered crystalline phases. The decagonal phase containing Cu showed more disordering compared to Al-Co-Ni alloys. The implication of the structural / microstructural changes due to Cu substitution in stable decagonal quasicrystal will be discussed.

\*    Corresponding Author: yadavtp22@rediffmail.com
                     Phone No.  +91-542-2368468, 2307307    fax: +91-542-2368468

## 1. Introduction

Quasicrystals are intriguing because they require one to reconsideration of all the basic concepts that have been developed for periodic crystal. After the discovery of icosahedral quasicrystals (IQC) by Shechtman et al. [1], many more alloy systems exhibiting various classes of quaiscrystal have been reported [2]. Among those the icosahedral and the decagonal phases have been studied most extensively. Icosahedral (I) phase is quasi-periodic in three dimension with a point group of m35, while the decagonal (D) phase is quasi-periodic in two and periodic in the other dimension, with point group 10/mmm [3]. The two stable class of two-dimensional quasicrystal having decagonal symmetry are reported in Al-Co-Ni and Al-Co-Cu ternary systems [4]. The stable decagonal phases with composition of $Al_{65}Co_{15}Cu_{20}$ and with different periodicities of $\approx$4,8,12 and 16Å were reported by He et al. [5] and confirmed by Tsai et al. [6]. Grushko et al. [7] studied the solidification behavior of D-$Al_{65}Co_{15}Cu_{20}$ and D-$Al_{65}Co_{20}Cu_{15}$ and found that the D-phase melts incongruently. The Al-Co-Cu phase diagram and stability of the D-phase were studied in detail by Grushko et al. [8]. The other stable decagonal quasicrystals in $Al_{70}Co_{15}Ni_{15}$ system have been investigated by Tsai et al. [9]. The analysis of X-ray diffraction results of $Al_{65}Co_{15}Cu_{20}$ has been shown to give rise to higher dimensional space group of $P10_5/mcm$, whereas $Al_{70}Co_{15}Ni_{15}$ belongs to the symmetry group of P10/mmm [10]. A subtle change in their diffraction characteristics between the decagonal phases of $Al_{70}Co_{15}Ni_{15}$ and $Al_{65}Co_{15}Cu_{20}$ has been observed in an electron diffraction study by Edagawa et al. [11].

The question concerning the stability of decagonal quasicrystal and the underlying stabilization mechanism are not yet fully understood. Grushko et al. [12] have investigated the transition between the periodic and quasi-periodic structures of decagonal Al-Co-Ni alloy of as-cast and annealed condition. The effect of the Si substitution for the conformation of the stability of Al-Cu-Co decagonal phase have been studied [13]. Recently the transition from decagonal to vacancy ordered phases (VOP) in $Al_{70}Co_{15-x}Cu_xNi_{15}$ alloy system have been investigated by Yadav et al. [14]. It is important to note that the possibility of the formation of different VOPs by changing the alloy composition has been discussed in details [15]. The substitution of elements for 3d transition metal atoms in $Al_{70-x}Co_{15}Cu_{x+y}Ni_{15-y}$ alloys leading to interesting structural variations has been reported by Pramanik et al. [16]. While deciding on the compositions of the alloys investigated here the stable decagonal phases namely the $Al_{70}Co_{15}Ni_{15}$ and $Al_{70}Co_{15}Cu_{15}$ are considered. In the present investigation, we have substituted Ni with Cu in the stable decagonal phase $Al_{70}Co_{15}Cu_xNi_{15-x}$ gradually up to the composition $Al_{70}Co_{15}Cu_{15}$. The phase evolution under rapid solidification condition has been investigated in detail.

## 2. Experimental Procedure

The alloy with nominal composition of $Al_{70}Cu_{15}Cu_xNi_{15-x}$ (x=0, 2.5, 5.0, 7.5, 10 and 15) were prepared in an argon atmosphere by melting high purity Al (99.96%), Co (99.98%), Ni (99.96%) and Cu (99.99%) using an RF induction furnace. Rapid solidification processing (RSP) of the as-cast alloys were conducted by melt spinning on to a copper wheel rotating at a speed of ~3600 rpm in an argon atmosphere. During the melt spinning, the entire apparatus was enclosed in a steel enclosure through which argon gas was made to flow continuously at pressure of 4.5MPa so as to prevent oxidation of the ribbons after ejection from the nozzle. The thickness and width of the RSP ribbon were found to be ~50 μm and ~ 0.75-1.0 mm, respectively. The structural characterization was done by employing a Philips PW-1710, X-ray diffractometer with a graphite monochromator and CuKα radiation. For characterizing the surface features scanning electron microscope JEOL 840A W3 with a Kevex Sigma-II energy depressive X-ray analyzer (EDX) analyzer was employed. The ribbons were thinned using an electrolyte of 92% ethanol and 8% Perchloric acid at $-14^{o}C$. The structural and micro-structural investigations of the sample were carried further out using a Philips EM-CM-12 electron microscope, operating at voltage of 100kV.

## 3. Results and Discussion

The surface morphology of the as cast $Al_{70}Co_{15}Ni_{15}$ and $Al_{70}Co_{15}Cu_{15}$ alloys show the aggregate of decagonal needles, which have been reported earlier by Yadav etal. [17]. The elongated needle structure has been characterized through TEM and identified as D-phase. The D-phase grows faster along the 10-fold direction compared to the other two directions. However the SEM micrographs of other alloys containing Cu do not show any regular or symmetric morphology like decagonal rods, rather it shows highly irregular shapes. This can be attributed to the effect of Cu in the quaternary alloy during solidification.

Powder X-ray diffraction (XRD) patterns of the $Al_{70}Co_{15}Cu_xNi_{15-x}$ (x = 0, 5, 10 and 15) alloys are shown in Fig. 1(a-d). After 5 at% of Cu substitution, (i.e. $Al_{70}Co_{15}Cu_5Ni_{10}$), there is a distinct peak broadening in x-ray diffraction pattern as exhibited by Fig.1 (b). After 10 at% Cu substitution (i.e. $Al_{70}Co_{15}Cu_{10}Ni_5$) in Fig. 1(c) both the decagonal and crystalline phases start evolving. It may be discerned from Fig.1 that 5 at% of Cu in $Al_{70}Co_{15}Cu_5Ni_{10}$ alloy has a tendency to destabilize the decagonal phase by increasing the substitutional disorder. When the Cu concentration is further increased, the precipitation of a crystalline phase similar to $Al_3Ni_2$ type vacancy ordered phase occur and this phase co-exist with the Al-Co-Cu type decagonal phase (Fig.1(c)). It is interesting to note that Mukhopadhyay et al. [18] have also observed an evolution of the microcrystalline B2 phases (related to $Al_3Ni_2$ phase) due to disordering in Al-Cu-Co-Si system while growing the single crystals of D phase.

Figure 2 shows the effect of Cu concentration on the full width at half maxima (FWHM) for the radial scan of the (102202) fundamental reflection (following the indexing scheme by Mukhopadhyay and Lord [19]) of $Al_{70}Co_{15}Cu_xNi_{15-x}$ (x =0-15) alloys. Here the FWHM is found increase with increasing Cu concentration up to 5 at % (i.e. in $Al_{70}Co_{15}Cu_5Ni_{10}$ alloy). On further addition of Cu, FWHM is found to decrease due to the decomposition of the disordered D phases to vacancy ordered phase and decagonal phase The increase in FWHM with Cu concentration can be understood in terms of introduction of homogenous strain field in the quasiperiodic and periodic decagonal planes. Such a strain field is characteristic of the quasicrystal structure just before the transformation / precipitations to the approximants phase [20]. It is evident from XRD that the precipitation of the crystalline phase occurs above the 5 at% of Cu from where the decrease in FWHM of the XRD peak has started. It should be mentioned here that from XRD, we have estimated the quasilattice parameters $a_R$ and c for $Al_{70}Co_{15}Cu_xNi_{15-x}$ (x =0,2.5,5,7.5,10 and 15).The value of $a_R$ for 5 at % Cu is minimum in our case .The general trend of variation of $a_R$ and c for these type of alloys is in conformity with the work reported by Steurer [21]. The values are shown in the table 1 along with values of e/a ratio.

The selected area electron diffraction patterns of $Al_{70}Co_{15}Ni_{15}$ and $Al_{70}Co_{15}Cu_5Ni_{10}$ RSP ribbons taken along the 10-fold direction are shown in Fig. 3. The diffraction spots in Fig.3 (a) are rather sharp and arranged at strictly fixed position. This also indicates that the phason strain causing the peak shift and splitting is not significant along the ten-fold axis of the decagonal phase. However ten-fold pattern obtained from $Al_{70}Co_{15}Cu_5Ni_{10}$ alloy show ample variation in the intensities and position in Fig.3 (b) and diffuse pentagonal arrangement of spots can be seen in the plotted circle. Fig. 4(a-f) shows the A2D and J type two fold patterns of these alloys respectively (following the notation of Yan etal. [22]). The selected area electron diffraction (SAD) patterns shows the presence of streaking/ diffuse row perpendicular to the periodic direction, as typified by the streaked diffraction rows shown in Fig. (b, e). It is very interesting to note that as we increase the Cu concentration in $Al_{70}Co_{15}Cu_xNi_{15-x}$ system, the streaking gets broader first and then gradually starts weakening. It remains very prominent at the composition of $Al_{70}Co_{15}Cu_5Ni_{10}$. This feature is evident in Fig.3 (b, e). The streaking can be explained in terms of the substitutinal disordering of Cu/Ni in quasipreodic planes stacked along 10-fold axis. Similar features have been reported by Tendeloo et al [23] and interpreted as due to disordering of pencil-like linear regions of the decagonal structure.

Figure 5 (a) shows the diffraction patterns from the alloy composition $Al_{70}Co_{15}Cu_{10}Ni_5$. The indexing of the SAD reveals the presence of super structure of $\tau_3$ (Vacancy –ordered) phase, it may be noticed that (111) reciprocal vector corresponding to B2 phase has been divided into six divisions (fig 5(a)). This suggests that modulation of B2 phase along [111] direction has becomes six times of the parent phase (i.e. B2 Phase). As we know, different $\tau$ phases are identified on the basis of the number of division made by Bragg peaks along [111] direction of B2 phase [24]. Therefore, the VOP in the present observation can be understood as the superstructure of $\tau_3$ by doubling its unit cell. It may be designated as $\tau_6$ phase. The details of the superstructure and its atomic ordering will be discussed elsewhere. Chattopadhyay et al. [24] first pointed out the possible link between the vacancy ordering and one-dimensional quasiperiodicity. Thus the formation of $\tau$ phase indicates the structural instability of D phase in the presence of Cu & Ni and the continuous transformation from D phase to crystalline approximant phase. The corresponding microstructure is shown in fig.5 (b). The gray and white area has been identified as superstructure of $\tau_3$ and decagonal phase respectively. It can be noted that this is first time we have observe superstructure of $\tau_3$ phase in Al-Cu-Co-Ni alloy system. It is interesting to notice that the microstructures exhibit good contact between the grain and these grains are somewhat faceted. However, as the Cu concentration increase to 15 at% it has been found (microstructure is not shown here) that the grains become almost entirely composed of the equiaxied D phase structures without the $\tau$ phases as mentioned above. This suggests that in absence of the Ni, the formation of $\tau$ phases appears to be difficult. Therefore, it can be understood that the formation of superstructure of $\tau_3$ phases requires the compositional variation, which ultimately controls the e/a ratio of the alloy and in turn stabilizes the $\tau$ phases. The stability of these phases at high temperature is being investigated through suitable heat treatment. This will help us to established phase equilibrium in these quaternary alloys and the role of e/a ratio on their stabilization.

## 4. Conclusions

After investigation of the $Al_{70}Co_{15}Cu_xNi_{15-x}$ alloys, the following conclusion can be drawn.

 The presence of Cu up to 5 at % causes maximum strain in quasi lattice structure and on further addition of Cu i.e. up to 10 at % results in strain relaxation by giving rise to the formation of crystalline, $\tau$ and decagonal phases. The disordering due to Cu in the quaternary alloy is clearly established. This is reflected in the morphology of the as-cast microstructure as well as the rapidly solidified microstructure and structure of the alloys investigated here. The evolution of the superstructure of $\tau_3$ phases, links the instability of decagonal phase containing Cu (10 at %) and Ni (5 at %). It is also clear that in the absence of Ni, the formation of $\tau$-phase is not favored due to the lack of appropriate e/a ratio, which appears crucial to stabilize the D and $\tau$ phases.

**Acknowledgement:**

The authors would like to thank Prof. S. Ranganathan, Prof. S.Lele, Prof. K. Chattopadhyay, Prof. D. H. Kim, Dr. M.A.Shaz and Shri Vijay Kumar (Scientific officer) for encouragement and interest in this work. One of the authors (T.P.Y.) acknowledges to the CSIR for the financial support of the Senior Research Fellowship, during which period a part of the work has been completed.

**Table:** The $a_R$, c and e/a value with Cu concentration in $Al_{70}Co_{15}Cu_xNi_{15-x}$

| Composition | $a_R(\text{Å})$ | $c(\text{Å})$ | e/a |
|---|---|---|---|
| $Al_{70}Co_{15}Ni_{15}$ | 4.00 | 4.00 | 1.73 |
| $Al_{70}Co_{15}Cu_{2.5}Ni_{12.5}$ | 3.96 | 4.00 | 1.79 |
| $Al_{70}Co_{15}Cu_5Ni_{10}$ | 3.80 | 4.03 | 1.83 |
| $Al_{70}Co_{15}Cu_{7.5}Ni_{7.5}$ | 3.81 | 4.03 | 1.87 |
| $Al_{70}Co_{15}Cu_{10}Ni_5$ | 3.82 | 4.03 | 1.91 |
| $Al_{70}Co_{15}Cu_{15}$ | 3.82 | 4.03 | 1.99 |

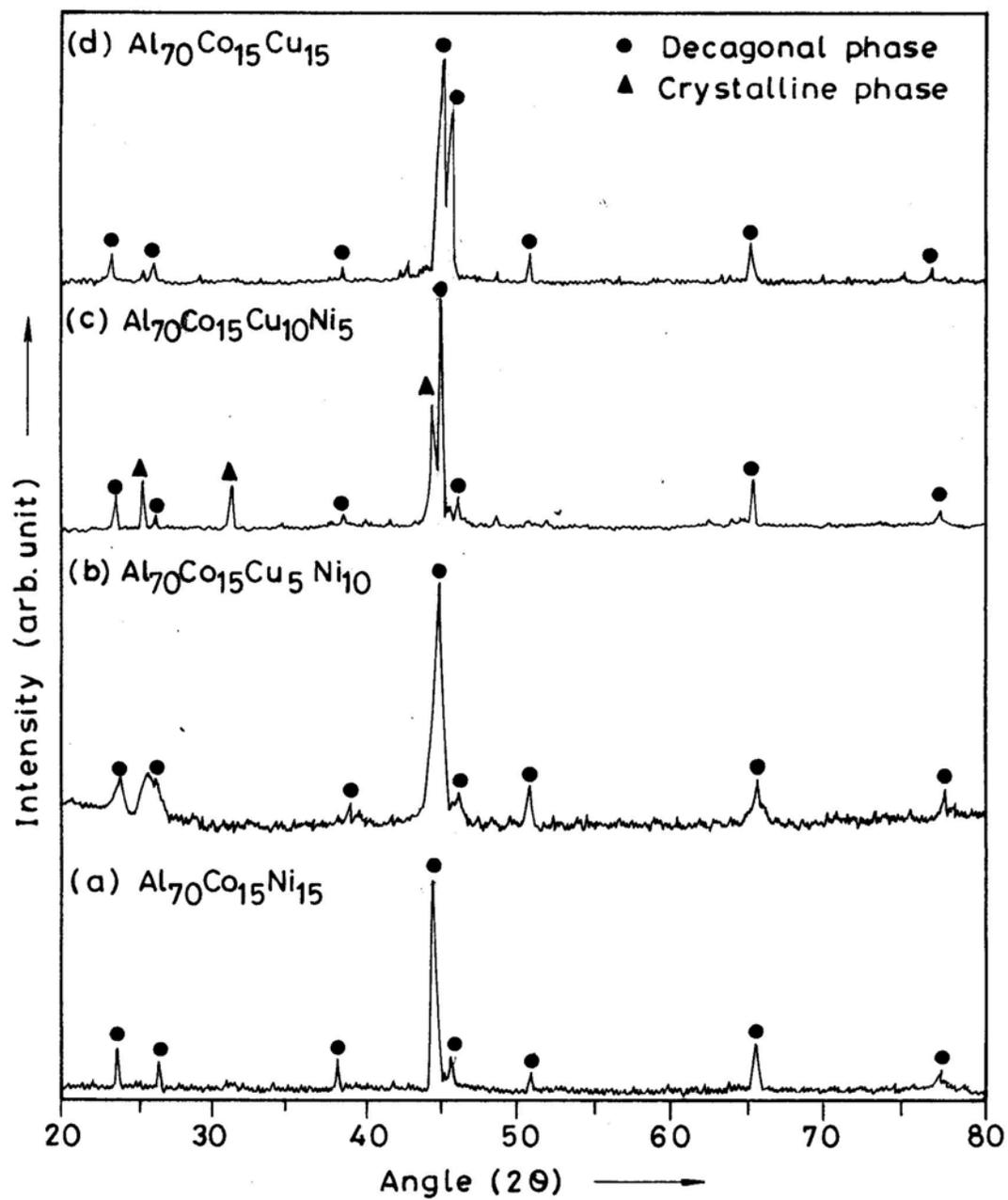

Fig.1   Powder X-ray diffraction (XRD) patterns of the $Al_{70}Co_{15}Cu_xNi_{15-x}$ (x = 0, 5, 10 and 15).

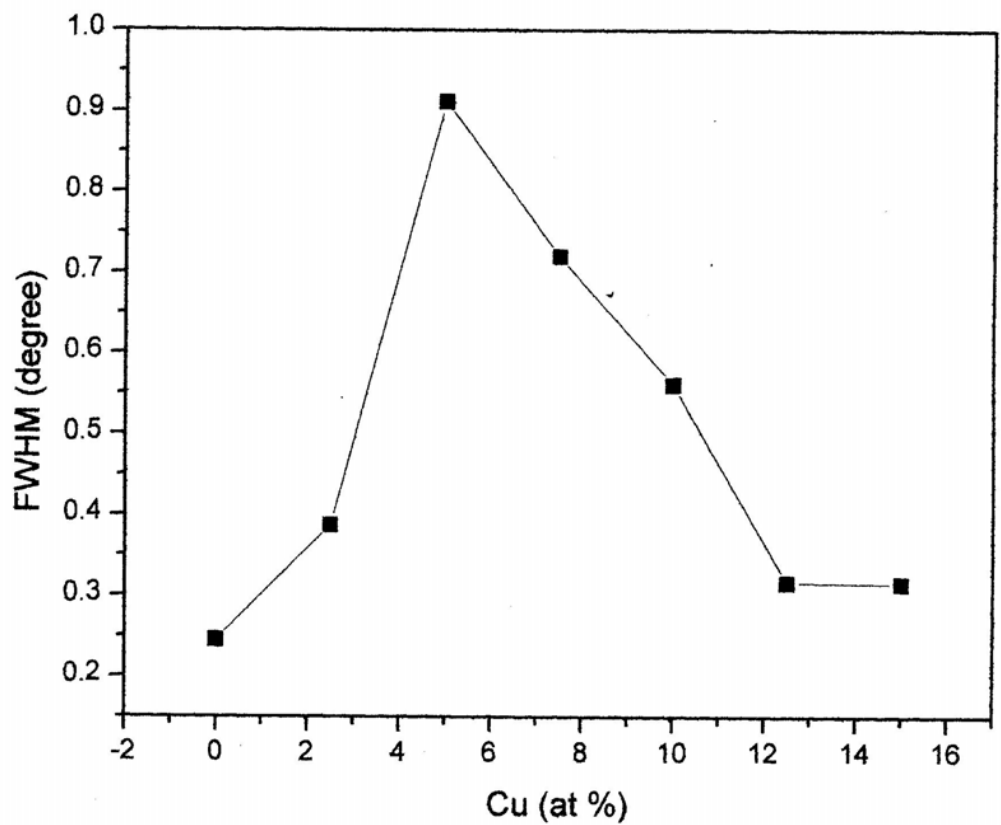

Fig.2 Cu concentration dependence of the FWHM for the radial scans of the (102202) fundamental reflection of $Al_{70}Co_{15}Cu_xNi_{15-x}$ (x =0-15) alloys composition.

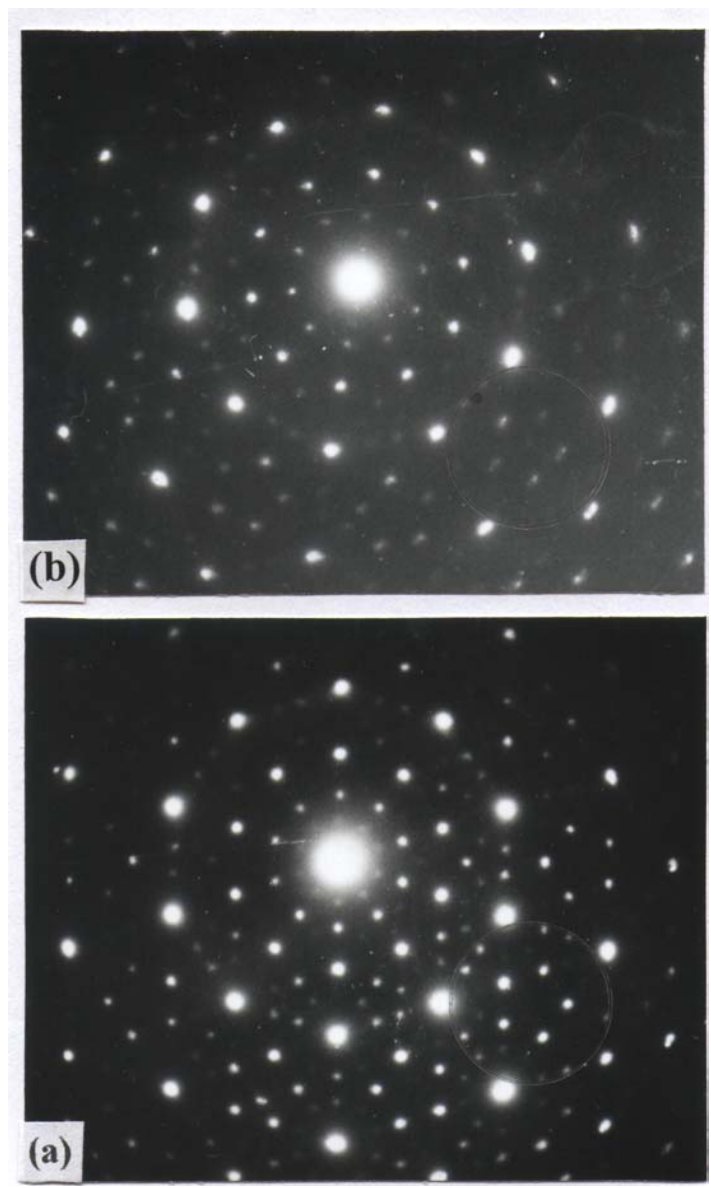

Fig.3. The selected area diffraction pattern of $Al_{70}Co_{15}Ni_{15}$ and $Al_{70}Co_{15}Cu_5Ni_{10}$ RSP ribbons taken along the 10 fold direction.

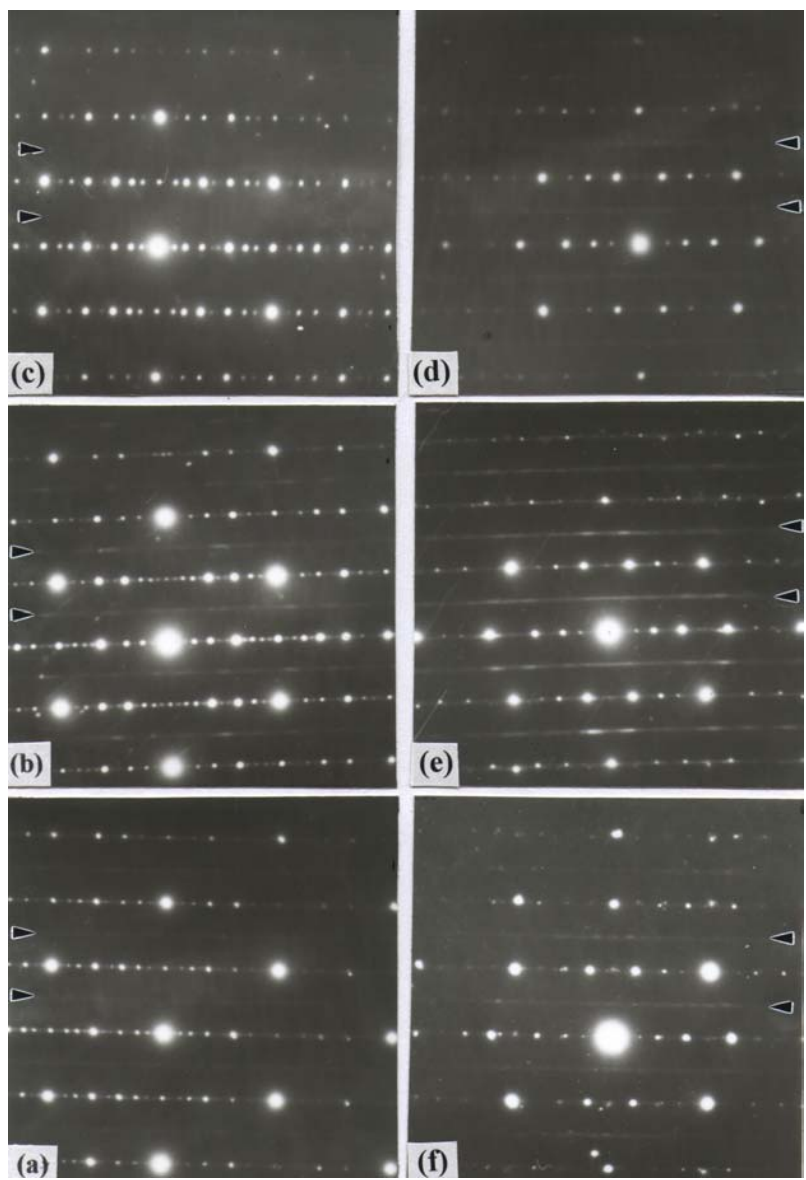

Fig.4    The selected area electron diffraction patterns of melt spun ribbon of the Al$_{70}$Cu$_x$Ni$_{15-x}$ (x = 0, 5, and 7.5) alloys. , (a-c) shows the A2D and (d-f ) shows J type two fold patterns of these alloys respectively

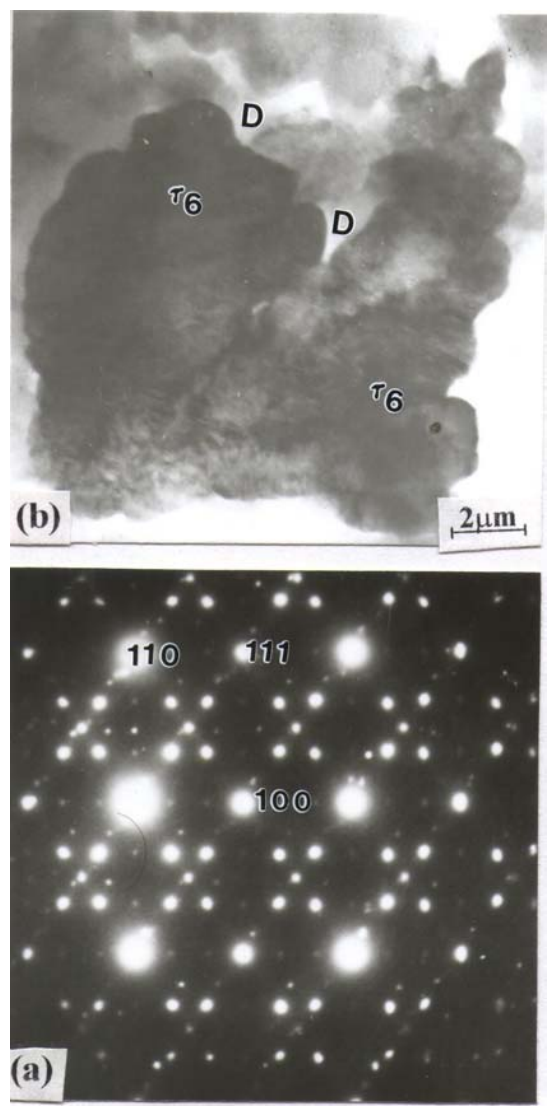

Fig. 5    (a) shows the diffraction patterns from the alloy composition $Al_{70}Co_{15}Cu_{10}Ni_5$

indicated the presence of $\tau$ phase  (b) shows the microstructure of superstructure of $\tau_3$ and D phase.